\documentclass{v15cosmo}

\def\be{\begin{equation}}
\def\ee{\end{equation}}
\def\bea{\begin{eqnarray}}
\def\eea{\end{eqnarray}}

\newcommand{\Photo}{\includegraphics[height=35mm]{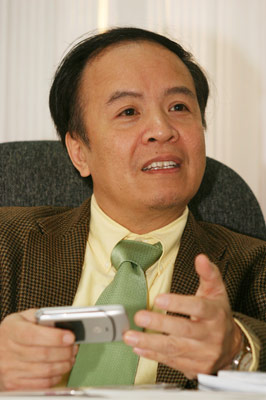}}

\begin{document}
\vspace*{4cm}
\title{ELECTROWEAK, STRONG INTERACTIONS AND HIGGS FIELDS AS COMPONENTS OF GRAVITY IN NONCOMMUTATIVE SPACETIME}

\author{NGUYEN AI VIET}

\address{Department of Physics, College of Natural Sciences 
and Information Technology Institute, \\
 Vietnam National University, E3 Building 144 Xuan Thuy, Cau Giay, Hanoi, Vietnam }

\maketitle\abstracts{
A new noncommutative spacetime of structure $ {\cal M}^4 \times Z_2 \times Z_2$ is proposed. The generalized Hilbert-Einstein action contains gravity, all known interactions and Higgs field. This theory can also provide a unified geometric framework for multigravity, which might explain the existence of dark matter and inflationary cosmology. In other words, high energy physics has laid out the crude shape of the new spacetime, while cosmology will shed light to the more details of it.}

\section{Noncommutative spacetime and nonabelian gauge fields}
\hspace*{12pt}Between 1921-1948, Kaluza, Klein and Thiry \cite{KK} had shown that the Hilbert-Einstein action in the spacetime extended with a circle ${\cal M}^4 \times S^1$ consisted of gravity, electromagnetism and a Brans-Dicke scalar. In 1968, R.Kerner \cite{Kerner} had generalized the Kaluza-Klein theory to include nonabelian gauge fields. Today, the multi-dimensional theories are studied widely as candidates of unified theories of interactions. However, these theories have a weakness of containing an infinite tower of massive fields leading to theoretical and observational obstacles.

In the 1980's, Connes had put forward the new concept of spacetime based on noncommutative geometry(NCG) \cite{Co}.In 1986, Connes and Lott \cite{CoLo} applied the idea to the spacetime extended by two discrete points ${\cal M}^4 \times Z_2$ and shown that Higgs fields emerged naturally in a gauge theory with a quartic potential. The most attractive feature of NCG with discrete extra dimensions is that it does not contains an infinte tower of massive fields.
 
In 1993, Chamsedine, Felder and Fr\"ohlich \cite{CFF} made the first attempt to generalize the Hilbert-Einstein action to NCG, leading to no new physical content. In 1994, G.Landi, N.A.Viet and K.C.Wali \cite{LVW} had overcome this no-go result and derived the zero mode sector of the Kaluza-Klein theory from the generalized Hilbert-Einstein's action. Viet and Wali \cite{VW1} have generalized this model further and obtained a full spectum consisting of bigravity, bivector and biscalar. In each pair, one field is massless and the other one is masive.

The incorporation of the nonabelian gauge fields in Viet-Wali's model is a not trivial task. Recently, Viet and Du \cite{VietDu} have successfully derived the nonabelian gauge interaction from the Hilbert-Einstein's action. However, it is possible to do so only in two following cases:

i. The gauge vector fields must be abelian on one sheet of spacetime and nonabelian on the other one. This is exactly the case of the electroweak gauge fields on the two copies of Connes-Lott's spacetime of chiral spinors.

ii. The gauge vector field must be the same on both copies of spacetime of chiral spinors. This is also the case of QCD of strong interaction.

So, NCG can "explain" the specific gauge symmetry structure of the Standard Model.
 
In this article, we propose a new noncommutative spacetime structure ${\cal M}^4 \times Z_2 \times Z_2$, which is the ordinary spacetime extended by two discrete extra dimension, each consists of two discrete points. In other words, this noncommutative spacetime consists of two copies of Connes-Lott's spacetime. The generalized  Hilbert-Einstein action in this new spacetime contains all the known interactions of Nature and the observed Higgs field. In a more general case, this theory can also lead to multigravity, which might be necessary to explain the dark matter and inflationary cosmology related observations. 
\section{A new model of noncommutative spacetime and gravity}
\hspace*{12pt}The noncommutative spacetime ${\cal M}^4 \times Z_2 \times Z_2$ is the usual four dimensional spinor manifold extended by two extra discrete dimensions given by two differential elements $DX^5$ and $DX^6$ in addition to the usual four dimensional ones $dx^\mu$. Each extra dimension consists of only two points. This structure can also be viewed as four sheeted space-time having a noncommutative differential structure with the following spectral triplet:

i) The Hilbert space ${\mathcal H}= {\mathcal H}^v \oplus {\cal H}^w $ which is a direct sum of two Hilbert spaces ${\mathcal H}^u = {\cal H}^u_L \oplus {\cal H}^u_R, u = v,w$, which are direct sums of the Hilbert spaces of left-handed and right-handed spinors. Thus the wave functions $ \Psi \in {\mathcal H}$  can be represented as follows 
\begin{equation}
    \Psi(x) =  \pmatrix{
    \Psi^v(x) \cr
    \Psi^w(x) \cr
    } ~~,~~ 
    \Psi^u(x) = \pmatrix{
        \psi^v_L(x) \cr
        \psi^w_R(x) \cr
        }  \in {\mathcal H}^u ~;~ u=v,w,
\end{equation}
where the functions $\psi^u_{L,R}(x) \in {\cal H}^u_{L,R}$ are defined on the 4-dimensional spin manifold ${\cal M}^4$.

ii) The algebra ${\cal A}={\cal A}^v \oplus {\cal A}^w ; {\cal}^u = {\cal A}^u_L \oplus {\cal A}^u_R$ contains the 0-form ${\cal F}$ 
\begin{equation}\label{0form}
   {\mathcal  F}(x) =  \pmatrix{
    F^v(x) & 0 \cr
    0 & F^w(x) \cr
    } ~,~ F^u(x) =  \pmatrix{
        f^u_L(x) & 0 \cr
        0 & f^u_R(x) \cr
        } \in {\cal A}^u,
\end{equation}
where $f^u_{L,R}(x)$ are real valued function operators defined on the ordinary spacetime ${\cal M}^4$ and acting on the Hilbert spaces ${\cal H}^u_{L,R}$.

iii) The Dirac operator ${\cal D} = \Gamma^P D_P = \Gamma^\mu \partial_\mu + \Gamma^5 D_5+ \Gamma^6 D_6, P=0,1,2,3,5,6 $ is defined as follows
\begin{eqnarray}\label{Dirac1}
{\cal D} &=& \pmatrix{
D & m_1 \theta_1 \cr
m_1 \theta_1 & D \cr
}~,~ D = \pmatrix{
d & m_2 \theta_2 \cr
m_2 \theta_2 & d \cr} ~,~ d = \gamma^\mu \partial_\mu \\ 
D_\mu {\cal F} &=& \pmatrix{
\partial_\mu F^v(x) & 0 \cr
0 & \partial_\mu F^w(x) \cr 
} ~,~
\partial_\mu F^u = \pmatrix{
\partial_\mu f^u_L(x) & 0 \cr
0 & \partial_\mu f^u_R(x) \cr 
} \\
D_{6} {\cal F} &=& m_1(F^v - F^w){\bf r} ~,~ D_5 F^u = m_2(f^u_L-f^u_R) {\bf r} ~,~ {\bf r} = \pmatrix{
1 & ~0\cr
0 & -1\cr
} 
\end{eqnarray}
where $\theta_1, \theta_2$ are Clifford elements $ \theta^2_1 = \theta^2_1=1$, $m_1, m_2$ are parameters with dimension of mass. 

The construction of noncommutative Riemannian geometry in the Cartan formulation is given in \cite{VW1} in a perfect parallelism with the ordinary one. Here we will use the following flat and curved indices to extend the 4 dimensions with 5-th and 6-th dimensions.
\begin{eqnarray}
E,F,G = A, \dot{6} ~,~ & A,B,C = a, \dot{5}&~,~ a,b,c =0,1,2,3 \\
P,Q,R = M,6 ~,~ &M,N,L = \mu, 5 &~,~\mu,\nu, \rho=0,1,2,3.
\end{eqnarray}

The construction of noncommutative Riemannian geometry \cite{VW1} is in a perfect parallelism with the ordinary one. The starting point is the locally flat reference frame, which is a linear transformation of the curvilinear one with the vielbein coefficients. For transparency, let us write down the vielbein in 4,5 and 6 dimensions as follows 
\begin{equation}
e^a = dx^\mu e^a_\mu(x) ~,~ E^A = DX^M E^A_M(x) ~,~{\cal E}^E = DX^P {\cal E}^E_P(x),
\end{equation}
where $e^a_\mu(x), E^A_M(x), {\cal E}^E_P$ are 4,5 and 6-dimensional vielbeins.
The Levi-Civita connection 1-forms $\Omega^\dagger_{EF} = - \Omega_{FE}$ are introduced as a direct generalization of the ordinary case. With a condition \cite{VW1}, which is a generalization of the torsion free condition one can determine the Levi-Civita connection 1-forms and hence the Ricci curvature tensor from the generalized Cartan structure equations    
\begin{eqnarray}
{\cal T}^E &= &  DE^E + E^F \Omega^E_F \label{Torsion}\\
{\cal R}^{EF} &=& D\Omega^{EF}+ \Omega^E_G \wedge \Omega^{GF} \label{Curv}
\end{eqnarray}
Then we can calculate the Ricci scalar curvature $R=\eta^{EG} \eta^{FH} R_{EFGH}$.

The construction of our model is carried out in two subsequent steps. First, we construct the 6-dimensional Ricci curvature with an ansatz containing one 5-dimensional gravity and two 5-dimensional vectors fields, where one is abelian and the other is nonabelian to use Viet-Du's results. Then
\begin{equation}
R_6 = R_5 - {1 \over 4}  G^{MN} G_{MN} = R_5 + {\cal L}_g(5)
\end{equation}
where $G_{MN}; M,N=0,1,2,3,5$ is the 5-dimensional covariant field streng tensor of the nonabelian $SU(2)\times U(1)$ gauge fields.

In the second step, the gravity sector is reduced further to 4-dimensional gravity  nonabelian gauge $SU(3)$ vector of strong interaction 
\begin{equation}
R_5 = R_4 - {1 \over 4} Tr H^{\mu \nu} H_{\mu \nu} ~~,~~ H_{\mu \nu} = \partial_\mu B_\nu - \partial_\nu B_\mu + i g_S [B_\mu, B_\nu],
\end{equation}
where $B_\mu = B^i_\mu(x) \lambda^i$ are the gluon field and $\lambda^i, i=1,..,8$ are the GellMann matrices.

Connes-Lott's procedure can be applied now to reduce the 5-dimensional gauge Lagrangian ${\cal L}_g(5)$ to the 4-dimensional electroweak gauge-Higgs sector of the Standard Model as follows
\begin{equation}
{\cal L}_g(5) = -{1 \over 4}( F^{\mu\nu} F_{\mu\nu}+ G^{\mu \nu} G_{\mu \nu}) + {1 \over 2} \nabla^\mu \bar{H} \nabla_\mu H + V(\bar{H}, H),  
\end{equation} 
where $H$ is a Higgs doublet, $\nabla_\mu$ is the gauge covariant derivative and $V(\bar{H}, H)$ is the usual quartic potential of the Higgs field.
\section{Multigravity in noncommutative spacetime}

\hspace*{12pt}In Section 2, we have presented the minimal ansatz to include the all known interactions and the Higgs fields. New cosmological observations might shed light to more detailed structure of the new commutative spacetime. In principles, in a more general case, our model can adopt up to 4 gravitational fields, one of those is massless while the other ones are massive. 

From theoretical points of view, this model can provide a geometric construction approach to the massive gravity, which has
recently attracted a lot of attention as a candidate theory of modified gravity \cite{DeRham}. From the viewpoints of modern cosmology, multigravity might give new explanations the existence of dark matter and inflationary  cosmology.
  
\section{Summary and discusions}
We have presented a new noncommutative spacetime ${\cal M}^4 \times Z_2 \times Z_2$, which can unify all the known interactions and Higgs field on a geometric foundation. This is very similar to the foundation of Einstein's general relativity.  This model unifies all forces in nature without resorting to infinite tower of massive fields.

In the most general case, this theory can contain more (but still a finite number) degrees of freedom, including four different massless and massive gravity fields, Brans-Dicke scalars and more gauge fields. The model can provide a geometric foundation to the theories of massive and modified gravity. The reality might be just a special case of the most general theory. The cosmological observations might help us to see more details of this theory.

There are some issues, at the moment we are not able to answer such as the physical meaning of the sixth dimension and the energy scale of this theory. It is worth to quote the following relation from the work by Viet and Du \cite{VietDu}
\begin{equation}
g = 8 m \sqrt{\pi G_N},
\end{equation}
where $g$ the weak coupling constant and $G_N$ is the Newton constant. This relation must hold when the theory becomes valid. One can speculate this might happen at an energy scale, which is million times lower that the Planck scale. That might be the case in some evolution stage of our universe after the Big Bang. All the above perspectives would merit more research.
\section*{Acknowledgments}
The discussions with Pham Tien Du, Do Van Thanh (College of Natural Sciences, VNU) and Nguyen Van Dat (ITI-VNU) are greatly appreciated. The author would also like to thank Jean Tran Thanh Van for the hospitality at Quy Nhon and supports. The work is partially supported by ITI-VNU and Department of Physics, College of Natural Sciences, VNU.

\end{document}